\documentclass[twocolumn,10pt,aps,prl]{revtex4-1}

\usepackage{amssymb, amsmath, amsthm}
\usepackage{graphicx}
\usepackage{units}

\begin{document}

\title{The cooperative Lamb shift in an atomic nanolayer}

\author{J. Keaveney$^{1}$}
\author{A. Sargsyan$^{2}$}
\author{U. Krohn$^{1}$}
\author{I. G. Hughes$^{1}$}
\author{D. Sarkisyan$^{2}$}
\author{C. S. Adams$^{1}$}
\email[Electronic address: ]{c.s.adams@durham.ac.uk}
\affiliation{$^{1}$ Department of Physics, Rochester Building, Durham University, South Road, Durham DH1 3LE, United Kingdom}

\affiliation{$^{2}$ Institute for Physical Research, National Academy of Sciences - Ashtarak 2, 0203, Armenia}

\date{\today}

\begin{abstract}
We present an experimental measurement of the cooperative Lamb shift and the Lorentz shift using an atomic nanolayer with tunable thickness and atomic density.
The cooperative Lamb shift arises due to the exchange of virtual photons between identical atoms. 
The interference between the forward and backward propagating virtual fields is confirmed by the thickness dependence of the shift which has a spatial frequency equal to $2k$, i.e. twice that of the optical field. The demonstration of cooperative interactions in an easily scalable system opens the door to a new domain for non-linear optics.
\end{abstract}

\maketitle

One of the more surprising aspects of quantum electrodynamics (QED) is that virtual processes give rise to real phenomena. For example, the Lamb shift arises from a modification of the transition frequency of an atom due to the emission and reabsorption of virtual photons. 
Similarly in cavity QED \cite{Goy1983,Jhe1987,Eschner2001} the reflection of the virtual field by a mirror modifies the absorptive and emissive properties of the atom. In a cooperative process such as superradiance the light-matter interaction is modified by the proximity of identical emitters. The dispersive counterpart of superradiance is known as the cooperative Lamb shift \cite{Friedberg1973} (also sometimes referred to as the collective or $\cal{N}$--atom Lamb shift \cite{Scully2010}). The cooperative Lamb shift and the cooperative decay rate (i.e. super- or subradiance) arise from the real and imaginary parts of the dipole--dipole interaction, respectively. 
Although superradiance has been investigated extensively \cite{Gross1982}, experimental studies of the cooperative Lamb shift are scarce.
Evidence for the shift is restricted to two particular cases, involving three-photon excitation in the limit of the thickness $\ell$ being much larger than the transition wavelength $\lambda$ in an atomic gas \cite{Garrett1990}, and X-ray scattering from Fe layers in a planar cavity \cite{Rohlsberger2010}, demonstrating the fundamental link between the cooperative shift and superradiance.
However, the full thickness dependence of the shift in a planar geometry with $\ell < \lambda$ predicted four decades ago \cite{Friedberg1973} has not been observed.

Here we present experimental measurements of the cooperative Lamb shift in a gaseous nanolayer of Rb atoms as a function of both density and length. The atoms are confined in a cell between two superpolished sapphire plates.  Similar nanolayers have been studied extensively over the last two decades, see e.g.  \cite{Briaudeau1998,Sarkisyan2001,Dutier2003,Dutier2003a,Sarkisyan2004,Fichet2007}.
We extend this work to the high density regime where dipole--dipole interactions dominate. In addition by building the effects of dipole--dipole interactions into a sophisticated model of the absortpion spectra we are able to extract the length depedence of the resonant shift and thereby verify that the spatial frequency of the cooperative Lamb shift is equal to twice that of the light field \cite{Friedberg1973}. We thus confirm the fundamental mechanism of the cooperative Lamb shift as the exchange of virtual photons.

The underlying mechanism of light scattering is the interference between the incident field and the local field produced by induced oscillatory dipoles.
In a medium with $N$ two level dipoles per unit volume the susceptibility for a weak field is given by the steady state solution to the optical Bloch equations (see e.g. \cite{Loudon1983})
\begin{eqnarray}
\chi &=& -\frac{N}{\epsilon_0\hbar}\frac{d^{2}}{\Delta+{\rm i}\gamma_{\rm ge}}~,
\end{eqnarray}
where $d$ is the transition dipole moment, $\gamma_{\rm ge}$ is the decay rate of the coherence between the ground and excited states and $\Delta$ is the detuning from resonance.
The response of an individual dipole is described in terms of the polarizability, 
\begin{eqnarray}
\alpha_{\rm p} &=& \frac{\chi}{4\pi N}=- \frac{1}{4\pi\epsilon_0 \hbar}\frac{d^2}{\Delta+{\rm i}\gamma_{\rm ge}}~.
\label{eq:alpha_p}
\end{eqnarray}
In a dense medium, the field produced by the dipoles modifies the optical response of each individual dipole. This modified response is found by adding the incident field to the dipolar field,  ${\cal E}_{\rm loc} ={\cal E}+{\cal P}/{3\epsilon_0}$~,
where ${\cal E}_{\rm loc}$ is known as the Lorentz local field \cite{Lorentz1909}. The susceptibility determines the bulk response ${\cal P}=\epsilon_0\chi{\cal E}$, whereas the polarizability determines the local response ${\cal P}=4\pi\epsilon_0N\alpha_{\rm p}{\cal E}_{\rm loc}$. Substituting for ${\cal E}$ and ${\cal P}$ we find a relation between the macroscopic variable $\chi$ and the single dipole parameter $\alpha_{\rm p}$ which is referred to as the Lorentz--Lorenz law \cite{Lorentz1909} 
\begin{eqnarray}
\chi&=&\frac{4\pi N\alpha_{\rm p}}{1-\frac{4}{3}\pi N\alpha_{\rm p}}~.
\end{eqnarray}
Substituting for $\alpha_{\rm p}$ we find
\begin{eqnarray}
\chi&=& - \frac{Nd^2/\epsilon_0\hbar}{\Delta+{\rm i}\gamma_{\rm ge}+Nd^2/3\epsilon_0\hbar}~,
\end{eqnarray}
and hence the first order correction due to dipole--dipole interactions is a shift in the resonance frequency known as the {\it Lorentz shift}
\begin{eqnarray}
\Delta_{\rm LL}&=& - \frac{Nd^2}{3\epsilon_0\hbar}~.
\end{eqnarray}
However, as discussed by Stephen \cite{Stephen1964} 
and Friedberg, Hartmann and Manaasah \cite{Friedberg1973} the pairwise dipole--dipole interaction also contains a radiation term. The complete pair potential for two dipoles, $V_{\rm dd}$, has the form  
\begin{align}
\begin{split}
V_{\rm dd} = \epsilon \left[  (1-{\rm i}kr)(3\cos^2\theta-1) +(kr)^{2}\sin^2\theta \right] {\rm e}^{{\rm i}kr}~,
\end{split}
\label{eq:dd}
\end{align}
where $\epsilon = -{3\hbar\Gamma}/{4(kr)^{3}}$, $r$ and $\theta$ are their separation and relative angle, respectively, and $\Gamma$ is the natural linewidth of the dipole transition with wavevector $k=2\pi/\lambda$.
The real and imaginary parts of $V_{\rm dd}$ give rise to a level splitting and a modification of the spontaneous lifetime (superradiance or subradiance), respectively \cite{Stephen1964,Lehmberg1970,Friedberg1973,Friedberg2010}.
While these effects have been demonstrated in experiments on two ions \cite{Devoe1996} and two molecules \cite{Hettich2002}, a key advantage in our experiment is  that we can easily vary the mean spacing $\langle r \rangle$ between atoms. By changing the temperature of the vapor between 20$^{\circ}$C and 350$^{\circ}$C we can smoothly vary the number density over 7 orders of magnitude. In doing so we move between two regimes: $Nk^{-3}\ll1$, $\langle r \rangle>\lambda$ where dipole--dipole interactions are negligible; and $Nk^{-3}\approx100$, $\langle r \rangle \sim \lambda/30$ where dipole--dipole interactions dominate.

For more than two dipoles the cooperative $\cal{N}$--atom shift and decay rate are given by a sum of the pairwise dipole--dipole interaction Eq. (\ref{eq:dd}) for all pairs. 
For the relatively simple case of an ensemble of dipoles confined within a thin plane of thickness $\ell$, the sum produces a shift \cite{Friedberg1973}
\begin{equation}
\Delta_{\rm dd}= -\vert\Delta_{\rm LL}\vert + \frac{3}{4}\vert\Delta_{\rm LL}\vert\left(1-\frac{\sin 2k\ell}{2k\ell}\right)~,
\label{eq:shift}
\end{equation}
where the first term is the Lorentz shift and the second term is the cooperative Lamb shift. 
There are two remarkable features of Eq.~(\ref{eq:shift}). First, the cooperative Lamb shift is a shift to higher energy. One can understand the opposite sign of the Lorentz shift and the cooperative Lamb shift from the pairwise potential, Eq.~(\ref{eq:dd}). 
For a thin slab where all the dipoles lie in the plane, all the dipoles oscillate in phase such that the dipole--dipole interaction reduces to the static case, which after averaging over all angles gives an attractive interaction resulting in the Lorentz shift to lower energy. 
As one moves out of the plane in the propagation direction the relative phase of the dipoles changes and at a separation of $\lambda/4$ the second dipole re--radiates a field that is $\pi$ out of phase with the source dipole. This switches the sign of the interaction giving rise to the cooperative Lamb shift to higher energies.  
The second interesting property of the shift is that it depends on twice the propagation phase $k\ell$ which arises due to the re-radiation by the second dipole \cite{Friedberg1973}. 
Finally we note that while superradiance requires excitation of the medium, the cooperative Lamb shift can be observed in the limit of weak excitation where there is negligible population of the excited state.

\begin{figure}[t]
\includegraphics[width=0.45\textwidth,angle=0]{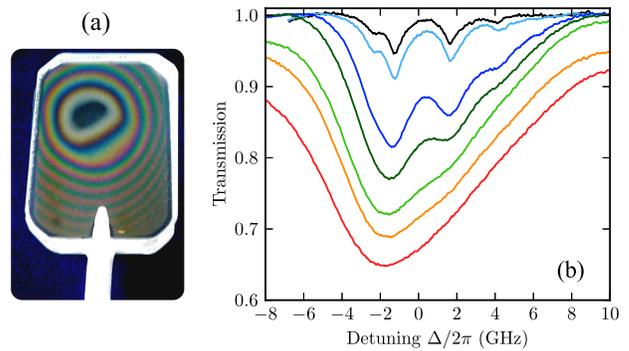}
\caption{(color online) Photograph of the nanocell and experimental data. 
(a) The Newton rings interference pattern can be observed on the windows of the cell due to the curvature of one of the windows, with a radius of curvature $> 100$~m. At the center of these rings the cell has its minimum thickness $\ell\sim 30$~nm, which increases to $\sim2$~$\mu$m near the stem at the bottom of the photo. 
(b) Transmission spectra for layer thickness $\ell = 90$~nm for measured Rb temperatures of \unit[190]{$^\circ$C} (black), \unit[207]{$^\circ$C} (light blue), \unit[250]{$^\circ$C} (blue), \unit[265]{$^\circ$C} (dark green), \unit[280]{$^\circ$C} (green), \unit[290]{$^\circ$C} (orange) and \unit[305]{$^\circ$C} (red).
The shift can be seen clearly as the density is increased. }
 \label{fig:shift}
\end{figure}
It is important to note that the shift $\Delta_{\rm dd}$ applies to a static medium. For a gaseous ensemble, atomic motion leads to collisions that also contribute a density dependent shift $\Delta_{\rm col}$ and broadening $\Gamma_{\rm self}$ of the resonance lines (see \cite{Weller2011a} and references therein), and thus the total shift for a thermal ensemble becomes
\begin{equation}
\Delta_{\rm tot}= \Delta_{\rm dd} + \Delta_{\rm col}~.
\label{eq:shift2}
\end{equation}
While evidence for density dependent shifts has been observed in experiments on selective reflection \cite{Maki1991}, it is important to measure $\Delta_{\rm tot}$ as a function of the length of the medium to separate the length independent collisonal shift $\Delta_{\rm col}$ \cite{Friedberg1973} from the length dependent cooperative Lamb shift. Below we present experimental data that allow that distinction to be made for the first time.

To measure the cooperative Lamb shift, we use a gaseous atomic nanolayer of Rb confined in a vapor cell with thickness $\ell < \lambda$. 
The cell is shown in  Fig.~1(a), and consists of a Rb reservoir and a window region,  where the Newton rings indicate the variation in the cell thickness from 30~nm at the centre to 2~$\mu$m near the bottom of the photograph. The wedge-shaped thickness profile arises due to the slight curvature of one of the windows (radius of curvature $R>100$~m). The local thickness at the position of the probe laser is measured at operational temperature using an interferometric method outlined in Ref.~\cite{Jahier2000}. 
The local surface roughness measured over an area of $1$~mm$^{2}$ is less than 3 nm, for any part of the window, and the focus of the beam is $\ll 1$~mm$^{2}$. 
The reservoir can be heated almost independently of the windows and its temperature determines the Rb number density, while the windows are kept $> 50^{\circ}$C hotter to prevent condensation of Rb vapor. By changing the temperature of the vapor between 20$^{\circ}$C and 350$^{\circ}$C we can vary the atomic density between the regimes $Nk^{-3}\ll1$ where dipole--dipole interactions are negligible and $Nk^{-3}\approx100$, where dipole--dipole interactions dominate. 

To determine the optical response of the medium we record transmission spectra as a narrowband laser is scanned across the D2 resonance in Rb at 780~nm. The light is reduced to a power $P\approx 100$~nW and focussed to a \unit[30]{$\mu$m} spot size inside the cell, leading to a local vapor length variation due to the wedge-shaped profile of less than 3~nm. The accuracy in determining the cell thickness is therefore limited by the surface flatness of the windows. Though the intensity of the light is greater than the conventional saturation intensity ($I_{\rm sat}\approx 1.7$~mW$/$cm$^{2}$ for the Rb D2 line), the extremely short length of the cell means that optical pumping is strongly suppressed.
The transmission is recorded on a photodiode, and a reference cell and Fabry-Perot interferometer are used to calibrate the laser frequency. Example experimental spectra for a thickness of $\ell=90$~nm are shown in Fig.~1(b), where the shift is clearly visible. The shift is extracted by fitting the observed spectra to a comprehensive model of the absolute transmission, based on a Marquardt-Levenberg method (see e.g. Ref. \cite{Hughes2010}). The model includes the effect of collisional broadening and has been shown to predict the absolute absorption of Rb vapor to better than 0.5\% \cite{Siddons2008b,Weller2011a}. To this we add the effects of Dicke narrowing \cite{Briaudeau1998}, where the Doppler effect is partially suppressed as a result of the short length scale; cavity effects \cite{Dutier2003a}, since the cell is a low-finesse etalon (with finesse $\mathcal{F} \sim 1$); and a single parameter which accounts for a ferquency shift of the whole spectrum.

\begin{figure}[t]
\includegraphics[width=0.48\textwidth]{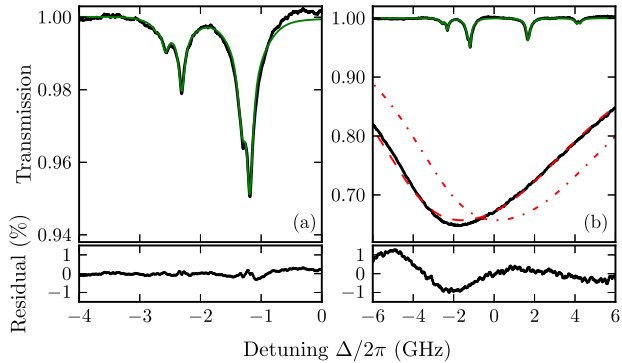}
\caption{(color online) Transmission spectra - experiment and theory. 
Transmission spectrum as a function of linear detuning for thickness (a) $\ell = 390$~nm, $Nk^{-3}\approx0.1$ ($T=130^{\circ}$C), and (b) $\ell = 90$ nm, $Nk^{-3}\approx50$ ($T=305^{\circ}$C). The black line is experimental data, while the solid green and dashed red lines are the fits to the model outlined in the main text. The dot-dashed red line in panel (b) is the theory without the line shift included. The residuals show the difference between experiment and theory. Zero on the detuning axis represents the weighted line centrer of the D2 line.}
\label{fig:2}
\end{figure}

Fig.~2 shows experimental data and the theoretical fit for two cases. Panel (a) shows a relatively large vapor thickness ($\ell = \lambda/2=390$~nm) with low atomic density where dipole--dipole interactions are negligible ($Nk^{-3}\approx 0.07$), and highlights the effects of Dicke narrowing at their most striking. 
Clearly visible are the individual excited state hyperfine components that are normally masked by Doppler broadening in a conventional cm-thickness cell.
In stark contrast to this, panel (b) shows the spectrum obtained in the dipole--dipole dominated regime ($Nk^{-3}\approx 50$) for a thickness $\ell=90$~nm. The individual hyperfine transitions are no longer resolved and there is a clear shift of the resonance to lower frequency. To illustrate this, we also plot the theoretical prediction with the line shift removed.
From fitting the data, the collisional broadening is found to be $\Gamma_{\rm self}=\sqrt{2} \cdot 2\pi N \Gamma k^{-3}=2\pi(1.0\times 10^{-7})N$~Hz cm$^{3}$ for thicknesses greater than $\lambda/4$ in agreement with previous work (see \cite{Weller2011a} and references therein). For thicknesses shorter than $\lambda/4$ we observe additional broadening that requires further investigation.
We also observe a van der Waals shift due to atom-surface interactions which we extract by fixing the density and varying the cell thickness (see also \cite{Fichet2007}), but even for the smallest thickness (90~nm) this is more than an order of magnitude smaller that the cooperative Lamb shift.

By comparing the experimental data with the theoretical prediction we extract the line shift as a function of number density and the thickness of the medium. 
In Fig.~3 we show the measured shift as a function of number density for two thicknesses, $\ell=90$~nm and $250$~nm. 
Hyperfine splitting gives rise to a different effective dipole for each transition in the spectrum, which at low densities shift independently. However, in the high density regime ($N>10^{16}$~cm$^{-3}$) dipole--dipole interactions dominate the lineshape and hyperfine splitting becomes negligible. We can then treat the line as a single s$_{1/2}\rightarrow$~p$_{3/2}$ transition which shifts linearly with density, as shown in Fig.~3.
\begin{figure}[t]
 \includegraphics[width=0.35\textwidth,angle=0]{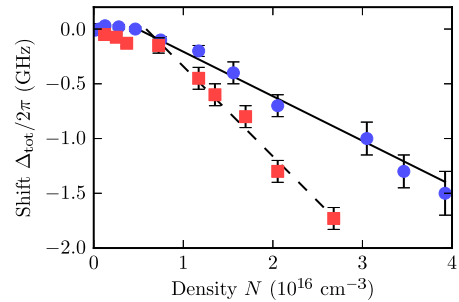}
\caption{(color online) Shift of resonance lines with density. Measured shift of resonance lines with density and fit to the linear, high density region for $\ell=90$ nm (red squares, dashed line) and $\ell=250$ nm (blue circles, solid line).}
 \label{fig:3}
\end{figure}
 We fit the gradient of the linear region to obtain the coefficient of the shift, and repeat these measurements for 13 thicknesses up to 600~nm. For thicknesses greater than 600~nm, the high optical depth of the sample impairs resolution of the line shift. We also observe anomalous behaviour around $\ell = 420$~nm, which may be due to the 5s-6p atomic resonance around 420~nm in Rb populated by the well-known energy pooling process \cite{Barbier1983}.

In Fig.~4 we plot the gradient of the line shift as a function of cell thickness. 
For the Rb D2 resonance, $\Delta_{\rm LL}/N= -2\pi \Gamma k^{-3}$, where we have used the relationship between the dipole moment for the s$_{1/2}\rightarrow {\rm p}_{3/2}$ transition and the spontaneous decay rate, $d=\sqrt{2/3} \langle L_{\rm e}=1 \vert er \vert L_{\rm g}=0\rangle$ (see Ref. \cite{Siddons2008b}).
We extract the collisional shift  by comparing the data to Eq.~(\ref{eq:shift2}) with $\Delta_{\rm col}$ the only free parameter. The amplitude and period of the oscillatory part are fully constrained by Eq.~(\ref{eq:shift}). We find the collisional shift to be $\Delta_{\rm col}/2\pi=(0.25\pm0.01)\times 10^{-7}$~Hz~cm$^{3}$, similar to previous measurements on potassium vapor \cite{Maki1991}. 
In this high density limit, the collisional shift is also independent of hyperfine splitting.  
The solid line is the prediction of Eq.~(\ref{eq:shift}), and the agreement between the measured shifts and the theoretical prediction is remarkable (the reduced $\chi^2$ for the data is 1.7). As well as measuring the thickness dependence of the cooperative Lamb shift, our data also provide a determination of the Lorentz shift which can only be measured in the limit of zero thickness.

\begin{figure}[t]
 \includegraphics[width=0.37\textwidth,angle=0]{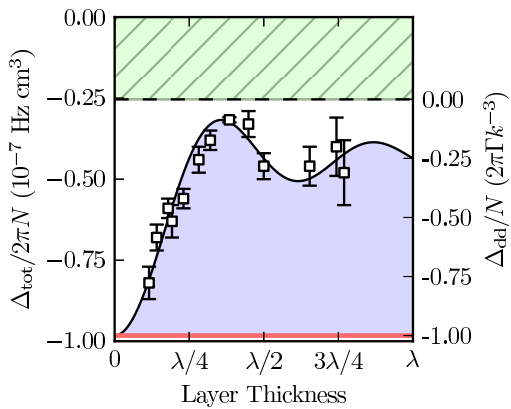}
 \caption{(color online) Experimental verification of the cooperative Lamb shift. The gradient of the shift $\Gamma_{\rm tot}/N$ is plotted against cell thickness $\ell$. The solid black line is Eq. (\ref{eq:shift2}) with $\Delta_{\rm col}/2\pi=-0.25\times10^{-7}$~Hz cm$^{3}$ and no other free parameters.
The coloured areas highlight the different contributions to $\Delta_{\rm tot}$; the Lorentz shift (red line), the cooperative Lamb shift (blue), and the collisional shift (green hatched). The alternate ordinate axis highlights the scaling between universal and experimental units.}
 \label{fig:4}
\end{figure}
The demonstration of the cooperative Lamb shift and coherent dipole--dipole interactions in media with thickness $\sim\lambda/4$ opens the door to a new domain for quantum optics, analgous to the strong dipole--dipole non-linearity in blockaded Rydberg systems \cite{Lukin2001,Pritchard2010}, that combines high bandwidth and high repetition rate with a simple optical set-up that is easily scalable. As the cooperative Lamb shift depends on the degree of exciation \cite{Friedberg1973}, exotic non-linear effects such as mirrorless bistability \cite{Fleischhauer1999,Bowden1979} are now accessible experimentally. In addition, given the fundamental link between the cooperative Lamb shift and superradiance, sub--quarterwave nanolayers offer an attractive system to study superradiance in the small volume limit. These topics will form the focus of future research.

We would like to thank M. P. A. Jones for stimulating discussions. We acknowledge financial support from EPSRC and Durham University.


%

\end{document}